\documentstyle[12pt]{article}
\title{Conformal Invariance and  Particle Aspects in
General Relativity}
\author{
 H. Salehi$^{a,b}$\thanks{e-mail: h-salehi@cc.sbu.ac.ir}\hspace{2mm},
 H. R. Sepangi$^{a}$\thanks{e-mail:hr-sepangi@cc.sbu.ac.ir}
 \hspace{1mm} and \hspace{0.5mm}F. Darabi$^{a}$\thanks{
e-mail: f-darabi@cc.sbu.ac.ir}
\\
\\
\\
$^{a}${\small Department of Physics, Shahid Beheshti University, Evin,
Tehran 19839,  Iran.}\\
$^{b}${\small Institute for Humanities
and  Cultural Studies, Tehran 14155-6419, Iran.}}
\begin{document}
\maketitle
\begin{abstract}
We study the breakdown of
conformal symmetry in a conformally invariant gravitational model. The 
symmetry breaking is introduced by defining a preferred
conformal frame in terms of the large scale characteristics of the universe.
In this context we show that a local change of the 
preferred conformal frame results in 
a Hamilton-Jacobi equation describing a particle with adjustable mass.
In this equation the dynamical characteristics of the particle 
substantially depends 
on the applied conformal factor and local geometry. 
Relevant interpretations of the results are also discussed.
\end{abstract}
\vspace{2cm}
\section{Introduction}

Conformal invariance has been playing a particularly important role in the
investigation of gravitational models ever since the emergence of such
theories. In a system which includes matter, it
is well known that conformal invariance requires the vanishing of the
trace of the stress tensor in the absence of
dimensional parameters.
In the presence of dimensional parameters, the conformal invariance 
can be established for a large class of theories \cite{Bekenstein}
if the dimensional parameters are conformally transformed according to their
dimensions. One general feature of conformally invariant theories is,
therefore, the presence of varying 
dimensional coupling constants. In particular, one can say that
the introduction of a constant dimensional parameter into
a conformally invariant theory breaks
the conformal invariance in the sense that a preferred conformal frame is 
singled out, namely that in which the dimensional parameter has
the assumed constant configuration. 
Thus the breakdown of conformal invariance 
may be established by introducing 
a constant dimensional parameter into the theory.
The determination of the corresponding preferred 
conformal frame depends on the nature
of the problem at hand. In a conformally invariant gravitational model
one usually considers the symmetry breaking 
as a cosmological effect. This would mean that one breaks the conformal
symmetry by defining a preferred conformal frame in terms of the large scale 
properties of cosmic matter distributed in a finite universe.
In this way, the breakdown of conformal symmetry was found to 
be a framework in which one can
look for the origin of the gravitational coupling of matter, 
both classical \cite{Deser}
and quantum \cite{Salehi}, at large cosmological scales.

The purpose of this paper is to show that 
the cosmological breakdown of conformal invariance 
in a conformally invariant gravitational model 
together with a local change of the
corresponding preferred conformal frame may be used to model a particle
concept in general relativity. 

The organisation of the paper is as follows:
We first study the breakdown of conformal symmetry in a 
conformally invariant gravitational model
and define
the resulting preferred conformal frame in terms of some cosmological
characteristics in close correspondence
to the work in \cite{Deser}. We then show that,
by a local change of the preferred conformal frame, 
it is possible to derive
a Hamilton-Jacobi type equation with adjustable mass.  
In our presentation
there is a substantial dynamical interplay between a 
particle in the ensemble and the applied conformal factor 
in a form which is similar to the dynamical effect of the
quantum potential on a particle 
in the context of the causal interpretation of relativistic quantum 
mechanics \cite{Holland}. 
This 
result seems to be interesting 
because it suggests that the emergence of 
quantal behaviour of matter may be a general feature of a 
conformal invariant gravitational model. 
We shall work with a metric having the signature (- + + +).

\section{Breakdown of conformal invariance}

In this section we briefly review the work in \cite{Deser}. Consider
the action functional
\begin{equation}
S[\phi]=\frac{1}{2} \int \!d^4 x \sqrt{-g} (g^{\mu \nu} \partial_{\mu} \phi
\partial_{\nu} \phi +\frac{1}{6} R \phi^2)
\label{1}
\end{equation}
which describes a system consisting of a real scalar field $\phi$ 
non-minimally coupled
to gravity, $R$ is the scalar curvature. Variations with respect
to $\phi$ and $g_{\mu \nu}$ lead to the equations
\begin{equation}
(\Box -\frac{1}{6} R)\phi=0
\label{2}
\end{equation}
\begin{equation}
G_{\mu \nu}=6\phi^{-2} \tau_{\mu \nu}(\phi)
\label{3}
\end{equation}
where $G_{\mu \nu}=R_{\mu \nu}-\frac{1}{2}g_{\mu \nu} R$ is the Einstein tensor
and
\begin{equation}
\tau_{\mu \nu}(\phi)= - [\nabla_\mu \phi \nabla_\nu \phi - 
\frac{1}{2}g_{\mu \nu}
\nabla_\alpha \phi \nabla^\alpha \phi] -
\frac{1}{6}(g_{\mu \nu}\Box -\nabla_\mu \nabla_\nu)\phi^2
\end{equation}
with $\nabla_\mu$ denoting the covariant derivative. Taking the trace
of (\ref{3}) gives
\begin{equation}
\phi(\Box -\frac{1}{6} R)\phi=0
\label{5}
\end{equation}
which is consistent with equation (\ref{2}). This is a consequence of the
conformal symmetry of action (\ref{1}) under the conformal transformations
\begin{equation}
\phi \rightarrow \bar{\phi}=\Omega^{-1}(x) \phi ,\:\:\:\:\:\:\:\:\:\:
g_{\mu \nu}\rightarrow \bar{g}_{\mu \nu}=\Omega^2 (x) g_{\mu \nu}
\label{6}
\end{equation}
where the conformal factor $\Omega(x)$ is an arbitrary, positive and smooth
function of space-time.
Adding a matter source $S_{m}$ independent 
of $\phi$ to the action (\ref{1}) in the form
\begin{equation}
S = S[\phi] + S_{m}
\label{7}
\end{equation}
yields the field equations
\begin{equation}
(\Box -\frac{1}{6} R)\phi=0
\label{8}
\end{equation}
\begin{equation}
G_{\mu \nu}=6\phi^{-2}[\tau_{\mu \nu}(\phi)+T_{\mu \nu}]
\label{9}
\end{equation}
where $T_{\mu \nu}$ is the matter energy-momentum tensor.
The following algebraic requirement
\begin{equation}
T_\mu ^{\mu}=0
\label{10}
\end{equation}
then emerges as a consequence of comparing the trace of (\ref{9})
with (\ref{8}). This implies that only traceless matter can couple
consistently  to such gravity models. 

We may break the conformal symmetry by adding a dimensional mass 
term $\frac{1}{2}\int\!d^4 x \sqrt{-g} \mu^2 \phi^2$, with 
$\mu$ being a constant mass parameter, to the action
(\ref{7}). This leads to
\begin{equation}
\mu^2 \phi^2=T_\mu ^{\mu}.
\label{12}
\end{equation}
Consequently, the field equations become
\begin{equation}
(\Box -\frac{1}{6} R-\mu^2)\phi=0
\label{13}
\end{equation}
\begin{equation}
G_{\mu \nu}-3\mu^2 g_{\mu \nu}=6\phi^{-2} [\tau_{\mu \nu}(\phi)+T_{\mu \nu}].
\label{14}
\end{equation}
Now the basic input is to consider 
the invariance breaking as a cosmological effect. This would mean that 
one may take $\mu^{-1}$ as the length scale characterizing the 
typical size of the universe $R_0$ and $T_\mu^\mu$ 
as the average density of the large scale distribution
of matter $ \sim M R_0^{-3}$, where $M$ is the mass of the universe.
This leads, as a consequence
of (\ref{12}) to the estimation of the 
constant background value of $\phi$
\begin{equation}
{\phi}^{-2} \sim R_0^{-2}(M/R_0^3)^{-1} \sim R_0/M \sim G
\label{15}
\end{equation}
where the well-known empirical cosmological relation $GM/R_0 \sim 1$
has been used. Inserting this background value of $\phi$ into the field
equations (\ref{13}), (\ref{14}) leads to the following set of Einstein equations
\begin{equation}
G_{\mu \nu}-3\mu^2 g_{\mu \nu}=6{\phi}^{-2}
T_{\mu \nu} \sim G T_{\mu \nu} 
\label{16}
\end{equation}
with a correct coupling constant $8\pi G$, and a
term $3\mu^2$ which appears as an effective cosmological constant $\Lambda$
of the order of $ R_0^{-2}$. 
The field equation (\ref{13}) for $\phi$  contains
no new information. 

We should emphasize that the 
invariance breaking explained above dealt with a broken
conformal invariance in the presence of a 
dimensional matter source which effectively appeared as
a cosmological
constant $\Lambda$. By implication, 
we get a preferred conformal frame $(\phi,g_{\mu\nu},\mu)$ for 
the gravitational variables, namely that in which $\phi^2\sim G^{-1}$, 
$\mu^2\sim\Lambda$ and $g_{\mu\nu}$ determined by the
field equations (\ref{16}). This preferred confromal frame has the remarkable
property that it incorporates a 
correct coupling of the cosmic matter to gravity. We shall call it 
the cosmological frame.

\section{Particle interpretation}

A key feature of any fundamental theory consistent
with a given
symmetry is that its breakdown would lead to effects which can have various
manifestations of physical importance. Therefore, in the case of conformal
symmetry, 
one would expect that the corresponding cosmological
invariance breaking, considered in the last section,
would have an important effect on the local structure of the underlying
theory. Here we would like to study one particular effect
which illustrates the local particle aspects. 

There are two chracteristic scales of squred mass in the cosmological frame 
$(\phi,g_{\mu\nu},\mu)$  which are subjected to the scale hierarchy 
\footnote{We use units in which $\hbar=c=1$. In these units $\phi$ has the
dimension of mass.}
\begin{eqnarray}
\mu^2 \ll \phi^2 \sim G^{-1}. \label{neq1}
\end{eqnarray}
In terms if this scale hierarchy, it is possible to single out a 
congruence of hypersurface orthogonal timelike curves, if we subject 
the corresponding generating vector field $\nabla_\mu S$ to the condition
\begin{eqnarray}
\nabla_\mu S\nabla^\mu S=-(\phi^2-\mu^2). \label{neq2}
\end{eqnarray}
We can observe that no known elementary particle can move along the
congruence defined by (\ref{neq2}) because of the large mass-scale defined 
by the right hand side of (\ref{neq2}). Actually, we know from (\ref{neq1})
that
$$\phi^2-\mu^2\sim\phi^2\sim G^{-1}=m_p^2$$
where $m_p$ is the Plank mass . 
The situation, however, changes if we consider a local change of the 
cosmological frame. To illustrate this point, let us write equation 
(\ref{neq2}) in a new frame $(\bar{\phi},\bar{g}_{\mu\nu},\bar{\mu})$ 
which is locally connected to the frame $(\phi,g_{\mu\nu},\mu)$ by a
conformal transformation (\ref{6}). Taking into account that $S$ as a 
dimensionless quantity does not transform under conformal invariance, we find
\begin{eqnarray}
\bar{\nabla}_\mu S\bar{\nabla}^\mu S=-(\bar{\phi}^2-\bar{\mu}^2)
\label{neq3}
\end{eqnarray}
with $\bar{\phi}=\Omega^{-1}\phi$, $\bar{\mu}=\Omega^{-1}\mu$ and
the bar quantities refering to the new frame. Now, 
from the field equation (\ref{13}) we find 
$$\bar{\mu}^2=\frac{\stackrel{-}{\Box}\bar{\phi}}{\bar{\phi}}-
\frac{1}{6}\bar{R}$$
which, if combined with equation (\ref{neq2}) leads to
\begin{eqnarray}
\bar{\nabla}_\mu S\bar{\nabla}^\mu S=-\bar{\phi}^2+\frac{\stackrel{-}
{\Box}\bar{\phi}}{\bar{\phi}}-\frac{1}{6}\bar{R}.
\label{neq4}
\end{eqnarray}
Now, if the new conformal frame $(\bar{\phi},\bar{g}_{\mu\nu},\bar{\mu})$ 
is taken to be subjected to 
\begin{eqnarray}
\bar{\nabla}_\mu S\bar{\nabla}^\mu \bar{\phi}=0 \label{neq5}
\end{eqnarray}
then equation (\ref{neq4}) can be interpreted as a Hamilton-Jacobi equation
for a particle with adjustable mass-scale $\sim\bar{\phi}^2$. The relation
(\ref{neq5}) ensures that this mass-scale will not change along the 
particle trajectory. In addition, we get a dynamical effect on the particle 
trajectories, reflected in the term $\frac{\stackrel{-}{\Box}\bar{\phi}}
{\bar{\phi}}-\frac{1}{6}\bar{R}$, which illustrates a modification of 
the particle mass due to the scalar curvature and the applied conformal
transformation. 
In summary, we observe that starting from the given preferred cosmological 
frame $(\phi,g_{\mu\nu},\mu)$, we may derive a particle concept
in terms of a corresponding local change of the conformal frame. 
In this approach 
no separate particle action needs to be introduced, 
that of a conformally invariant gravitational model suffices. The particle 
aspects emerge from an internal condition connecting
the local properties of a time-like congruence of curves associated
with a characteristic scale hierarchy in the cosmological frame 
with particle properties in a new conformal frame.
This observation emphasizes
that general relativity, if suitably formulated as a conformal invariant 
field theory, does not 
ascribe any special significance to a separate particle action.
We should, however, note that there is a certain limitation on the 
applicability of the particle concept. Actually, in a universe in which 
$\phi^2$ is smaller than $\mu^2$, the above particle concept 
does not apply, for the mass scale of (\ref{neq2})
becomes tachionic.
This limitation, however, does not seem to be a weakness of our particle
concept because  the condition $\phi^2<\mu^2$ describes a 
universe of trans-Plankian size.

\section{Concluding remarks}

In this paper we have shown that the cosmological 
breakdown of conformal symmetry in a
conformally invariant gravitational model together
with a local change of the corresponding preferred conformal frame 
leads to a picture consistent
with a particle concept. 
This picture may be considered as a manifestation of Mach's 
principle in that the particle concept emerges as a local effect emerging
from large scale cosmological consideration.\\
We emphasize the similarity of the term
$\frac{\stackrel{-}{\Box}\bar{\phi}}{\bar{\phi}}$ on the right hand side of 
(\ref{neq4}) with the quantum potential term in the context of the causal
interpretation of relativistic quantum mechanics [3]. 
This similarity merits attention because it provides 
an indication for a possible
geometrization of quantal behaviour of relativistic particles in the framework
of a conformal invariant gravitational model.

%\newpage

\end{document}